\PassOptionsToPackage{unicode}{hyperref}
\PassOptionsToPackage{hyphens}{url}
\PassOptionsToPackage{dvipsnames,svgnames,x11names}{xcolor}
\documentclass[
  12pt]{article}

\usepackage{amsmath,amssymb,tikz}
\usepackage{iftex}
\ifPDFTeX
  \usepackage[T1]{fontenc}
  \usepackage[utf8]{inputenc}
  \usepackage{textcomp} 
\else 
  \usepackage{unicode-math}
  \defaultfontfeatures{Scale=MatchLowercase}
  \defaultfontfeatures[\rmfamily]{Ligatures=TeX,Scale=1}
\fi
\usepackage{lmodern}
\ifPDFTeX\else  
\fi
\IfFileExists{upquote.sty}{\usepackage{upquote}}{}
\IfFileExists{microtype.sty}{
  \usepackage[]{microtype}
  \UseMicrotypeSet[protrusion]{basicmath} 
}{}
\makeatletter
\@ifundefined{KOMAClassName}{
  \IfFileExists{parskip.sty}{%
    \usepackage{parskip}
  }{
    \setlength{\parindent}{0pt}
    \setlength{\parskip}{6pt plus 2pt minus 1pt}}
}{
  \KOMAoptions{parskip=half}}
\makeatother
\usepackage{xcolor}
\makeatletter
\ifx\paragraph\undefined\else
  \let\oldparagraph\paragraph
  \renewcommand{\paragraph}{
    \@ifstar
      \xxxParagraphStar
      \xxxParagraphNoStar
  }
  \newcommand{\xxxParagraphStar}[1]{\oldparagraph*{#1}\mbox{}}
  \newcommand{\xxxParagraphNoStar}[1]{\oldparagraph{#1}\mbox{}}
\fi
\ifx\subparagraph\undefined\else
  \let\oldsubparagraph\subparagraph
  \renewcommand{\subparagraph}{
    \@ifstar
      \xxxSubParagraphStar
      \xxxSubParagraphNoStar
  }
  \newcommand{\xxxSubParagraphStar}[1]{\oldsubparagraph*{#1}\mbox{}}
  \newcommand{\xxxSubParagraphNoStar}[1]{\oldsubparagraph{#1}\mbox{}}
\fi
\makeatother

\usepackage{longtable,booktabs,array}
\usepackage{calc} 
\usepackage{etoolbox}
\makeatletter
\patchcmd\longtable{\par}{\if@noskipsec\mbox{}\fi\par}{}{}
\makeatother
\IfFileExists{footnotehyper.sty}{\usepackage{footnotehyper}}{\usepackage{footnote}}
\makesavenoteenv{longtable}
\usepackage{graphicx}
\makeatletter
\def\maxwidth{\ifdim\Gin@nat@width>\linewidth\linewidth\else\Gin@nat@width\fi}
\def\maxheight{\ifdim\Gin@nat@height>\textheight\textheight\else\Gin@nat@height\fi}
\makeatother
\setkeys{Gin}{width=\maxwidth,height=\maxheight,keepaspectratio}
\makeatletter
\def\fps@figure{htbp}
\makeatother

\makeatletter
\@ifpackageloaded{caption}{}{\usepackage{caption}}
\AtBeginDocument{%
\ifdefined\contentsname
  \renewcommand*\contentsname{Table of contents}
\else
  \newcommand\contentsname{Table of contents}
\fi
\ifdefined\listfigurename
  \renewcommand*\listfigurename{List of Figures}
\else
  \newcommand\listfigurename{List of Figures}
\fi
\ifdefined\listtablename
  \renewcommand*\listtablename{List of Tables}
\else
  \newcommand\listtablename{List of Tables}
\fi
\ifdefined\figurename
  \renewcommand*\figurename{Figure}
\else
  \newcommand\figurename{Figure}
\fi
\ifdefined\tablename
  \renewcommand*\tablename{Table}
\else
  \newcommand\tablename{Table}
\fi
}
\@ifpackageloaded{float}{}{\usepackage{float}}
\floatstyle{ruled}
\@ifundefined{c@chapter}{\newfloat{codelisting}{h}{lop}}{\newfloat{codelisting}{h}{lop}[chapter]}
\floatname{codelisting}{Listing}

\makeatother
\makeatletter
\@ifpackageloaded{caption}{}{\usepackage{caption}}
\@ifpackageloaded{subcaption}{}{\usepackage{subcaption}}
\makeatother

\ifLuaTeX
  \usepackage{selnolig}  
\fi
\usepackage[]{natbib}
\usepackage{bookmark}

\IfFileExists{xurl.sty}{\usepackage{xurl}}{} 
\hypersetup{
  pdftitle={Explicational statistics: research for an AI age},
  pdfauthor={Sihai Dave Zhao},
  pdfkeywords={chronic social defeat, dimension reduction, manifold learning, single-cell RNA-seq},
  colorlinks=true,
  linkcolor={blue},
  filecolor={Maroon},
  citecolor={Blue},
  urlcolor={Blue},
  pdfcreator={LaTeX via pandoc}}

\newcommand{\anon}{1}

\usetikzlibrary{positioning}

\begin{document}

\def\spacingset#1{\renewcommand{\baselinestretch}%
{#1}\small\normalsize} \spacingset{1}


\if1\anon
{
  \title{\bf Explicational statistics: research for an AI age}
  \author{Sihai Dave Zhao\thanks{
    The author gratefully acknowledges support from the National Institutes of Health (R01AT013189).\hspace{.2cm}\\
  Department of Statistics and Carl R. Woese Institute for Genomic Biology, University of Illinois}}
  \maketitle
} \fi

\if0\anon
{
  \bigskip
  \bigskip
  \bigskip
  \begin{center}
    {\LARGE\bf Title}
\end{center}
  \medskip
} \fi

\bigskip
\begin{abstract}
  AI is dramatically changing the statistical research landscape. I argue that production of a distinct category of statistical knowledge, which I refer to as explicational statistics, is actually becoming more valuable as AI increasingly automates other parts of the research process. I define explicational statistics, contextualize it in terms of other forms of statistical knowledge, and explain why I think the rise of AI makes it more valuable. I demonstrate my framework by first developing a new dimension reduction method, Ambient Embedding Projection, that can simultaneously visualize the intrinsic geometry of a data manifold along with how that manifold is embedded into ambient space. I then use my new method to conceive of a new statistical property of a collection of manifolds, related to metric distortion, that in single-cell biology can capture the ensemble behavior of cell populations. I apply these ideas to single-nuclei RNA-seq data from the brains of mice from different behavioral conditions to reveal potential cellular mechanisms of chronic social defeat stress.
\end{abstract}

\noindent%
\vfill

\newpage
\spacingset{1.8} 

\section{Introduction}\label{sec-intro}

Statisticians, like most other communities, are trying to determine how best to evolve alongside AI \citep{donoho2026rebuilding, Lin2025Statistics, sosa2026statistics, wright2026importance}. Much of the discussion has focused on the effects of AI on statistical practice, training, and the visibility of the discipline. For example, one strategy has been to expand the scope of statistics to engage more deeply with other discplines \citep{he2025statistics, he2026statistics}, such as through the further development of data science.

My goal in this paper is to analyze a different aspect of the field's evolution, namely the changing nature of statistical research. The central challenge is that AI appears to have made it cheap to solve statistics research problems. AI can design and perform analyses, implement methodology, and prove theorems with unrivaled breadth and speed \citep{castelvecchi2026openai}. Of course, solving problems with AI is not yet trivial. It does not work well in all cases and its effective use currently still requires skill and creativity \citep{dobriban2025solving, jin2026counting, sun2026statistical}. Nevertheless, solving research problems is much easier than before, and AI tools will only improve from here.

This is causing a sea change in the value structure of the statistical research enterprise. There appear to be two trends. The first is to lean in to execution \citep{zhu2026statisticians}: we should take advantage of this new technology to solve as many interesting and important problems as we can! The second trend is to assign greater value to research tasks that AI tools cannot yet do well. One such task is problem formulation \citep{cheng2025barbarians, sun2026statistical}, where the work is in choosing an important problem and specifying it to such a precise extent that AI can help find its solution. Another task is verification and understanding \citep{sun2026statistical, tao2026mathematics}, especially in proof-based research, for, much like the tree that falls alone in a forest, if a theorem is proved but no one understands the proof, does the proof really exist?

So far, these discussions have centered on the \textit{processes} of statistical knowledge production.  However, AI does not only affect the research process, it is also changing the value structures underlying which \textit{categories} of statistical knowledge are being produced.

In this paper, I delineate a distinct category of statistical knowledge that I believe is becoming increasingly valuable in this AI age, and which may be a promising direction for growth for the discipline. Briefly, this type of knowledge comprises connections between theoretical constructs of some substantive field and quantitative constructs of mathematical statistics, thus providing \textit{explication} in the sense of \citet{carnap1945two}. Similar ideas have been previously discussed in statistics: in his 1986 presidential address to the Royal Statistical Society, \citet{nelder1986statistics} defined statistics as ``being primarily concerned with the theory and practice of the \textit{matching} of theory to data'' (emphasis mine). However, it seems that this category of statistical knowledge has not yet been carefully defined or deeply discussed.

In Section \ref{sec-framework}, I more precisely articulate this category of knowledge, and in Section \ref{sec-ai} I propose that the production of this type of knowledge is now especially valuable. In Section \ref{sec-illustration}, I give a concrete illustration of how this type of research might look, by developing a new dimension reduction method motivated by the analysis of single-cell RNA-seq data. I made heavy use of AI to review literature, learn new material, and implement the methodology in the illustrative example, and I document my usage throughout this paper.

\section{Explicational statistics}\label{sec-framework}
\subsection{Definition}\label{sec-definition}

Statistics is often viewed as a mathematical formalization of induction from quantitative data. 
Under this view, statistical knowledge would then pertain to understanding the foundational principles enabling this induction, the degree to which it is possible, how it should be designed and performed, and the uncertain nature of its conclusions.

However, this view of statistics seems incomplete. To see this, imagine that a group of researchers are interested in understanding American demographics, and suppose that they know the ages of every single American. However, suppose they are not aware of the discipline of statistics.

On the one hand, since they have access to the population distribution function $F(x)$, there is no need for induction, and thus no need for statistics if it were defined as above. On the other hand, it is difficult, without specialized training, to obtain substantive understanding from such a complicated object as $F(x)$. If they wanted to, the researchers could of course calculate $\int x dF(x)$, or $\int x^2 dF(x) - \{\int x dF(x) \}^2$, but they would probably not think to do so. Even if they were told that these quantities were useful, they may not immediately see why, as for example $\int x dF(x)$ does not \textit{prima facie} seem like it corresponds to anything meaningful.

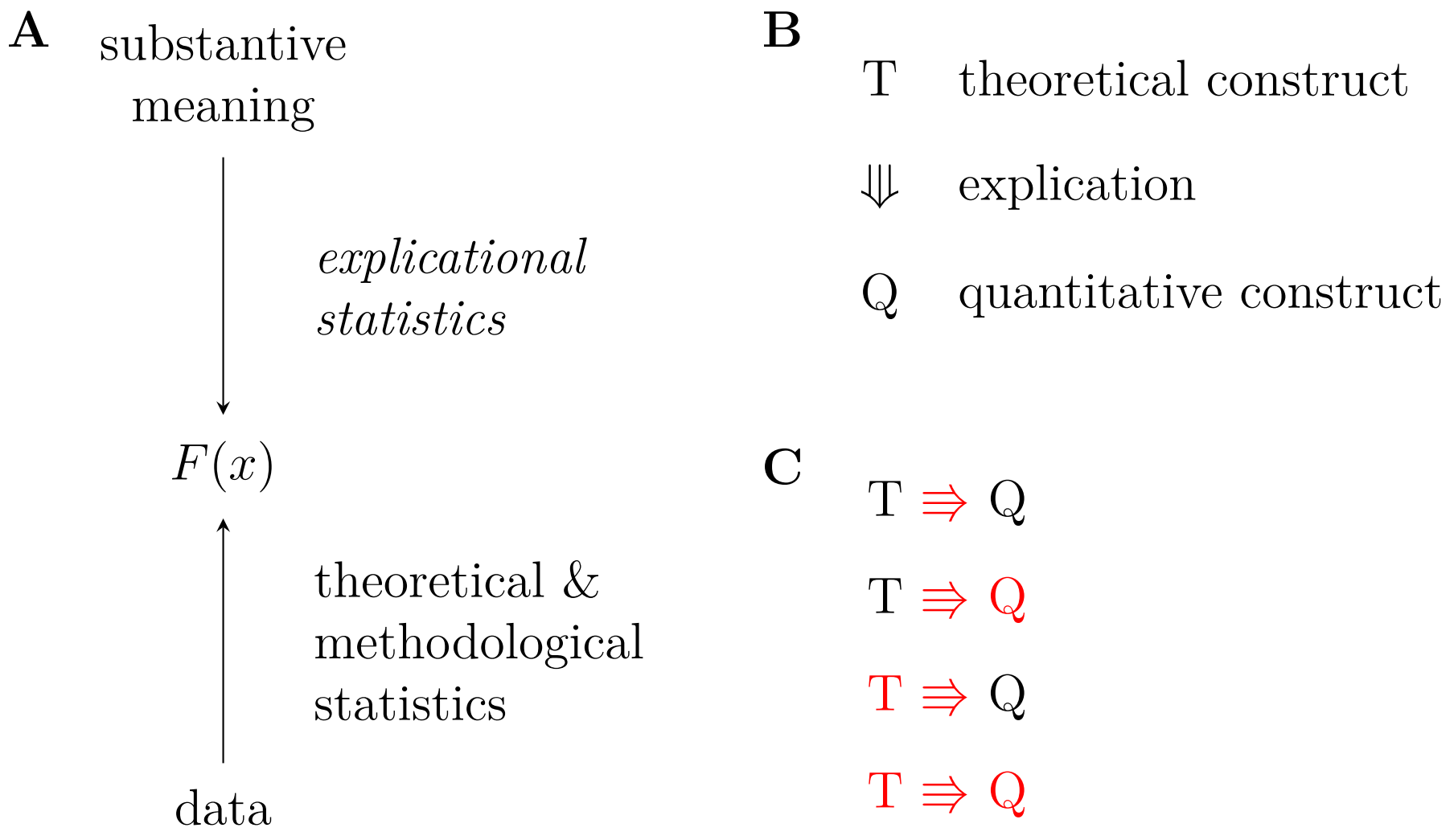
\begin{figure}
  \centering{
    
\begin{tikzpicture}[
  panel/.style={font=\bfseries},
  lbl/.style={align=left, anchor=west}
]
 
\node[panel] at (-0.6,0.4) {A};
 
\node[align=center, font=\fontsize{12}{8}\selectfont] (theory) at (1,0)    {substantive\\meaning};
\node (F)      at (1,-3.2) {$F(x)$};
\node (data)   at (1,-6)   {data};
 
\draw[-stealth, shorten <=2pt, shorten >=2pt] (theory) -- (F)
  node[midway, right=0.6cm, lbl, font=\fontsize{12}{8}\selectfont] {\textit{explicational}\\\textit{statistics}};
 
\draw[-stealth, shorten <=2pt, shorten >=2pt] (data) -- (F)
  node[midway, right=0.6cm, lbl, font=\fontsize{12}{8}\selectfont] {theoretical \&\\methodological\\statistics};
 
\node[panel] at (5.6,0.4) {B};
 
\node at (6.4,0)    {T};
\node[lbl] at (6.9,0)    {theoretical construct};
 
\node at (6.4,-0.9) {\rotatebox{-90}{$\Rrightarrow$}};
\node[lbl] at (6.9,-0.9) {explication};
 
\node at (6.4,-1.8) {Q};
\node[lbl] at (6.9,-1.8) {quantitative construct};
 
\node[panel] at (5.6,-3.2) {C};
 
\node[anchor=west] at (6.15,-3.5) {T\ {\color{red}$\Rrightarrow$} Q};
\node[anchor=west] at (6.15,-4.3) {T\ {\color{red}$\Rrightarrow$ Q}};
\node[anchor=west] at (6.15,-5.1) {{\color{red}T\ $\Rrightarrow$} Q};
\node[anchor=west] at (6.15,-5.9) {\color{red}T\ $\Rrightarrow$ Q};

\end{tikzpicture}
}
\caption{\label{fig-explication}A) Explicational statistics connects population-level constructs, like the distribution function $F(x)$, to substantive meaning. B) This connection is composed of three components. C) New explicational statistics can arise in different ways depending on which of these components is new (in red).}

\end{figure}%

This thought experiment reveals another type of statistical knowledge beyond the mathematization of induction: statistics has developed a quantitative language for describing populations that is connected to substantive meaning and understanding (Figure \ref{fig-explication}A). The mean and variance are simple examples of this language. For a more complex example, if the researchers were interested in how sharply age cohorts change in the American population, $\int \{ d^3F(x)/ dx^3 \}^2 dx$ might be a useful construct. The field of statistics thus not only describes how to do induction, it also defines the meaningful population-level constructs that become the targets of induction.

I propose that these definitions be delineated as a distinct category of statistical knowledge, which might be referred to as \textit{explicational statistics}. Explication was introduced by \citet{carnap1945two} in work on the foundations of probability theory and can be defined as follows \citep{Carnap1950}:
\begin{quote}
  By an \textit{explication} we understand the transformation of an inexact prescientific concept, the \textit{explicandum}, into an exact concept, the \textit{explicatum}.
\end{quote}
Thus a piece of explicational statistical knowledge can be defined as a triplet (Figure \ref{fig-explication}B), consisting of 1) a substantive theoretical construct, or explicandum, 2) a quantitative construct, or explicatum, and 3) the connection between them, or explication. In the mean example above, the explicandum would be the theoretical construct of ``typical'' age, the explicatum would be $\int x dF(x)$, and the explication would be the precise matching of the concept of typicalness to the single number that is closest to all observations in the sense of squared error. In roughness example, the explicandum would be the theoretical construct of ``sharpness'' of age cohorts, the explicatum would be $\int \{d^3 F(x) / dx^3\}^2 dx$, and the explication would be the precise matching of the concept of sharpness to the total amount of curvature in the density function.

What distinguishes explication in statistics from, say, explication in other areas of applied math is that the quantitative constructs of statistics describe features whose values can vary across populations. \citet{fisher1922mathematical} wrote that ``the object of statistical methods is the reduction of data'', which ``by its mere bulk incapable of entering the mind.'' Explicational statistics is the invention of useful reductions that explicate substantive meaning. I used Claude to identify Carnap's concept as the right philosophical expression of my ideas.

\subsection{Research}\label{sec-research}


Because it contains three elements, explicational knowledge might be created in multiple ways (Figure \ref{fig-explication}C). First, the explication may be new even if the theoretical and quantitative constructs that it connects are not. For example, that genes in a cell regulate each others' functions in a complex network of interactions is a well-known construct in molecular biology, and the correlation matrix is a well-known construct in statistics. The identification of the latter as a useful explication of the former, however, was a new idea in computational biology. This led to the idea of gene coexpression networks \citep{langfelder2008wgcna, markowetz2007inferring}, now a mainstay of the field.

Alternatively, a known theoretical construct may lead to a new explication and quantitative construct. For example, the known construct of geometric similarity between objects in space motivated the definition of the Gromov-Wasserstein distance metric \citep{memoli2011gromov}. Similarly, a known quantitative construct can lead to a new explication and theoretical construct, which happens frequently in physics \citep{Wigner1960}.

\begin{figure}
\centering{
\includegraphics[width=2.5in,height=\textheight]{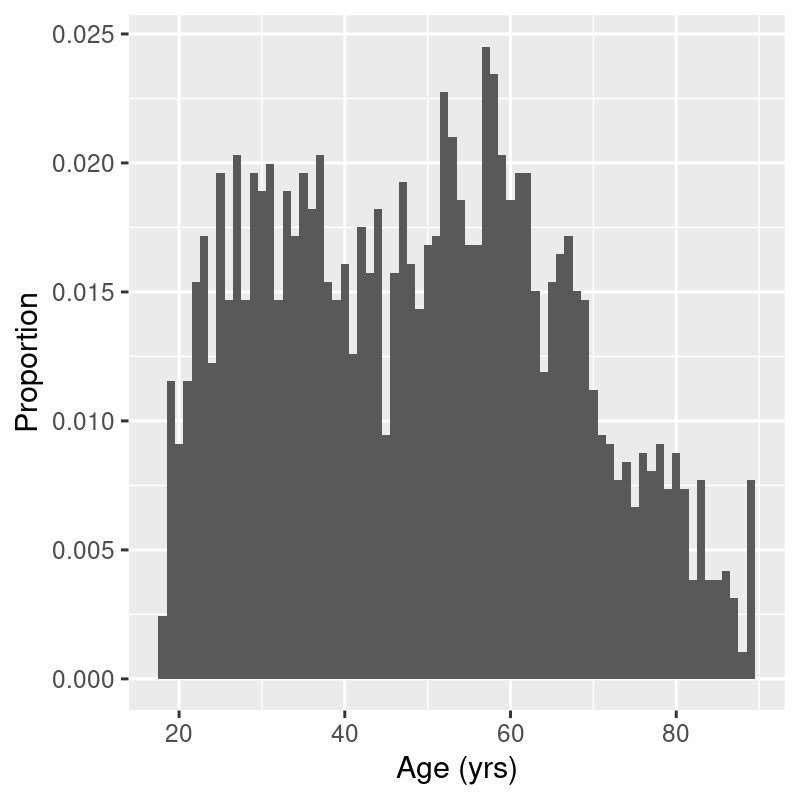}
}
\caption{\label{fig-age}Distribution of ages from the 2016 General Social Survey \citep{davern2025gss}, which can illustrate how all three elements of explicational statistical knowledge might be produced simultaneously.}

\end{figure}%

In general, however, producing new explicational statistics is very difficult because of the breadth of knowledge required. Without being aware of what theoretical constructs might be of interest to researchers, it is difficult to know what quantitative constructs might be meaningful to create. Conversely, without familiarity with the quantitative language of mathematical statistics, it is not clear what theoretical constructs can even be conceived for a study. In my experience, this is the main intellectual challenge of interdisciplinary collaboration.

Thus the most common, and most powerful, form of explicational knowledge production occurs when all three elements are co-constructed simultaneously. Consider again the thought experiment of the statistically naive researchers studying American demographics. Now suppose they conducted their study in 2016 but know nothing about the history of the country. They might produce new explicational statistics as follows:
\begin{enumerate}
\item Initially, the researchers likely would not even know what substantive questions they can ask that can be answered by the population age distribution $F(x)$.
\item If they happened to visualize $F(x)$, they would find something like Figure \ref{fig-age} (in reality this is the sample age distribution of respondents of the 2016 General Social Survey). 
\item This might motivate them to create the concept of bimodality (quantitative).
\item Bimodality might motivate them to ask the question of why there appear to be two distinct cohorts of Americans born at different times (theoretical).
\item To make precise this notion, the researchers might invent the two-component Gaussian mixture model (quantitative). Fitting the mixture model to the population would reveal that the older cohort has an average age of around 64 years.
\item Some investigation of American history circa 1952 might lead the researchers to identify the Baby Boomers (theoretical).
\end{enumerate}
Iteration between theoretical and quantitative constructs thus leads to new ideas in both regimes and new connections between them, and is similar to the account of \citet{bailyn1977research} of quantitative research as a cognitive process. In Section \ref{sec-illustration} I give a real illustration of this process by proposing new theoretical and quantitative constructs for single-cell transcriptomics.

\subsection{Relationships to other forms of statistical knowledge}

Explicational statistics is related to applied, methodological, and theoretical forms of statistical knowledge, but does not fit neatly within those traditional categories. The latter two mostly presuppose the existence of a statistical explication and are more concerned with induction of the explicatum. In these contexts, explication is most similar to the idea of model specification \citep{Lehmann1990ModelSpecification, Zhao2025What}. However, model specification usually refers to making simplifying assumptions about the data generating process. The goal of explicational statistics is less about modeling the data and more about modeling a substantive theoretical construct. This can take the form of a probability model, but also might take the form of a functional of a probability model, for example a mean or variance.

Like explicational statistics, applied statistics also involves interaction between statistics and another substantive discipline. Applied statistics is a broad term. It is sometimes thought of as the practice of statistics, in which case it describes procedural knowledge, while explicational statistics can be considered propositional knowledge. At the same time, it is almost never the case that statisticians are asked merely to implement to well-formulated analyses. Otherwise, if a quantitative explication has already been established for a desired theoretical construct, AI can already, or will soon be able to, do the job.

Explicational statistics is maybe most closely related to interdisciplinary fields like neurostatistics, computational biology, and quantitative social science. Motivated by the social sciences, \citet{Gelman_Hill_Vehtari_2020} note that matching measurements to underlying substantive constructs is a major goal of statistics. This is one example of explication, which applies beyond measurement. In sociology, \citet{lundberg2021your} discuss the importance of defining the estimand of interest in a quantitative study, which ``connects statistical evidence to theory''. The process of defining the estimand is exactly explication.

A reasonable question is why explicational statistics should be considered a part of statistics, rather than a part of the substantive field whose theoretical constructs are explicated. First, explicational statistics can produce new quantitative constructs, which are clearly contributions to mathematical statistics. Second, an explication associated with an existing quantitative construct can be considered as a property of that construct, and since the construct is a statistical object, the explication can be considered one as well.

But this question does point to the unique nature of explications as intersectional objects. Explicational statistics are triplets (Figure \ref{fig-explication}B), and therefore only exist in the interaction between a substantive field and statistics. This is why they belong most naturally to the interdisciplines mentioned above. At the same time, the more generally applicable an explicational statistic is, the more it might be ascribed to pure statistics. For example, the idea of a ``typical value'' is a theoretical construct that arises in nearly every substantive field. This is why the explication of the typical value as the mean is considered a solidly statistical object and not an interdisciplinary one.

As mentioned in Section \ref{sec-intro}, there is consensus within the statistical community that engagement with other disciplines is important for the future of the field. Explicational statistics describes one possible mode of engagement, and articulates the nature of the underlying intellectual contributions.

\section{Explicational statistics and AI}\label{sec-ai}

I believe research in explicational statistics will become dramatically more valuable as AI capabilities accelerate, and may be a fruitful direction for statisticians to consider. One main reason is that explication is foundational. It is the \textit{sine qua non} of data analysis. The explication defines the inferential goal for AI tools (and for traditional statistics and machine learning methods).

That being said, it is also true that AI can formulate its own problems. For example, there are many recent academic and private efforts to develop agents for automated scientific discovery \citep{feng2026agentic}. However, I think there are structural reasons why AI struggles with sophisticated explicational work.

First, as described in the co-construction example in Section \ref{sec-research}, the most powerful kind of explicational knowledge production occurs when we do not know exactly what questions we would like to ask. The creative challenge is in finding the right expression of our initial vague intuitions about what might be important in a problem. AI can certainly help us brainstorm theoretical and quantitative constructs that may be useful, but the selection of the ``right'' pair depends on what we want, and when we cannot even exactly articulate what we want, it will likely be hard for AI to automate the process.

Second, high-level explications change the way we understand the world. \citet{gowers2008princeton} describes two kinds of quantitative definitions. One kind simply serves as shorthand. Other definitions, however, ``make you think in new ways once you know them.'' Gowers points to the derivative of a function as one example, without which it would be impossible to even conceive of certain problems; ``the real breakthrough is the concept itself.'' These new conceptual structures are by definition not part of the data that AI is trained on, and so far it has not been clearly demonstrated that AI can develop new ways to conceive of the world on its own \citep{ball2026could}.

Finally, good explicational statistics requires value judgements. There is a vast number of theoretical constructs that can be explicated, quantitative constructs that can be defined, and problems that can be formulated. AI may be able to automatically enumerate all of them, but which ones are most valuable to pursue depends on the value systems of the substantive and statistical disciplines. These may not exactly match the value system encoded into an AI model. Furthermore, some constructs are difficult to value before they exist. For example, AI itself was considered a dying field in the late 1980s. Truly disruptive innovations are sometimes in direct opposition to existing value structures \citep{christensen1997innovators}, and it is not clear whether AI can decide when to buck the values it is trained on.

\section{Illustration}\label{sec-illustration}

\subsection{Motivation}

\begin{figure}
\centering{
\includegraphics[width=4in,height=\textheight]{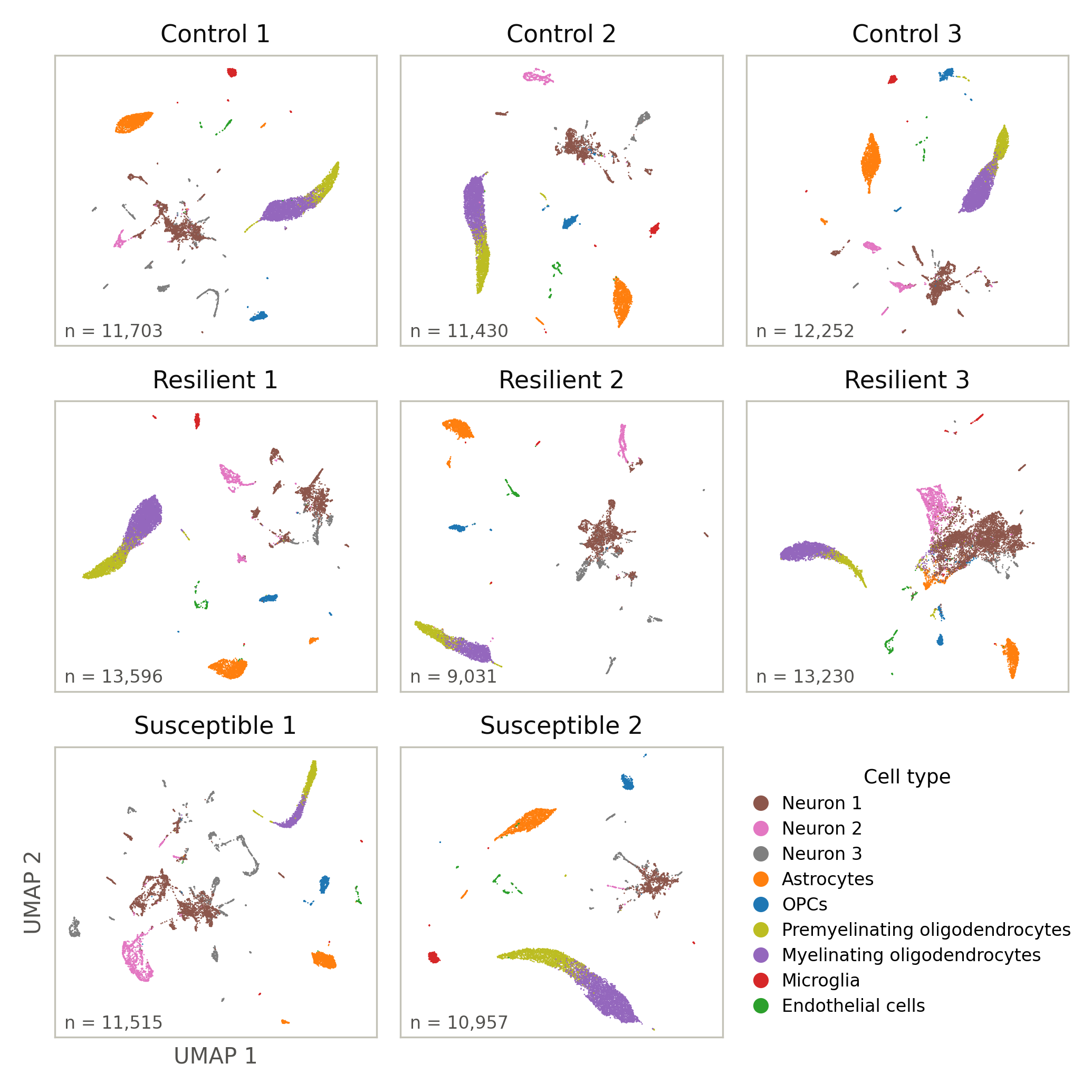}
}
\caption{\label{fig-vta}UMAP plots of single-nuclei RNA-seq data from \citet{minerva2026stress} of the ventral tegmental areas (VTAs) of male mice in a chronic social defeat experiment. Different numbers indicate different biological replicates, each consisting of pooled nuclei from two animals of the same condition. I used Claude Code to generate the figure.}

\end{figure}%

Here I illustrate how explicational statistics can be produced. My motivation comes from the study of the genomic basis of social behavior. Single-cell sequencing of brain tissue has been used extensively for this purpose \citep[e.g.,][]{johnson2023cellular, traniello2023single}. I will analyze data from \citet{minerva2026stress}, who collected single-nuclei RNA-seq (snRNA-seq) data from the ventral tegmental areas of mice who had undergone chronic social defeat (Figure \ref{fig-vta} plots data from the male mice). Briefly, they placed mice in a cage with an aggressor mouse, separated by a perforated barrier, for a few minutes for each of 10 days and then scored the mice based on a social interaction test to be either resilient or susceptible \citep{golden2011standardized}. The biological question was to understand the transcriptional mechanisms underlying the different responses to the stressful experience. I used Claude Science to download the data from the Gene Expression Omnibus (GSE303538) and perform quality control according to the original publication. 

The standard analysis approach is to call cell types, find cell type-specific differentially expressed genes, determine enriched pathways and gene regulatory networks, and perhaps test for cell-cell communication \citep{luecken2019current}. However, a rich line of work in sociogenomics \citep{robinson2002sociogenomics, robinson2005sociogenomics} has shown that a social stimulus causes such large global changes in the brain that its effects may be better understood in terms of shifts in the brain’s overall cellular transcriptomic state, rather than in terms of individual cells, cell types, or circuits. 

Thus, ensemble properties of single-cell transcriptomes, rather than properties of cells or cell types in the ensemble, may better capture the complexity of molecular mechanisms of social behavior. My goal was therefore to find a suitable ensemble property and quantitative explication, with the constraint that they still be interpretable at the level of single cells and cell types, which are the preferred abstractions in biology \citep{brenner2010sequences}.

\subsection{Previous work}

Ensemble properties of populations of cells have been studied by summarizing gene expression from the entire collection of cells from a replicate in some way and comparing the summarized representation across conditions \citep{boyeau2025deep, liu2026learning, wang2024gloscope}. However, these representations do not have convenient reductionist interpretations in terms of cells or cell types.

A different approach is suggested by the neuronal electrophysiology literature, where behavior is sometimes thought to be encoded in the ensemble firing activity of neuronal populations. This activity can be modeled as lying on a low-dimensional ``neural manifold'' that has been shown to differ between different behavioral conditions \citep{langdon2023unifying, perich2025neural}. This approach is also atractive because of the importance of the manifold hypothesis in machine learning and AI \citep{fefferman2016testing}.

In single-cell transcriptomics, cells are also viewed as lying on a low-dimensional manifold embedded in gene expression space \citep{moon2018manifold}. In the brain, this might be thought of as a ``neurogenomic manifold'', analogous to the electrophysiological neural manifold. Notions of the shape and geometry of this manifold may therefore be useful explications of the global transcriptomic changes that accompany social behavior.

Using Claude to survey the literature found existing work that has suggested that intrinsic geometric properties of these cell manifolds, such as curvature, may be related to biological phenomena like developmental trajectories \citep{bhaskar2022diffusion, huynh2024scgeom, sritharan2021curvature}. However, most of this work focuses on cell-level rather than ensemble-level properties, and also ignores the way the manifold is embedded into gene expression space, which is biologically meaningful.

\subsection{Proposed Ambient Embedding Projection method}

To find a more suitable quantitative construct and global property of the neurogenomic manifold, I realized it would be useful to simultaneously visualize the manifold's intrinisic structure as well as how it is embedded in ambient space. There does not appear to exist a method capable of doing this, though methods that visualze only instrinsic geometry abound \citep{maaten2008visualizing, mcinnes2018umap, tenenbaum2000global}, and PCA or multidimensional scaling can be used to visualize only extrinsic geometry.

I therefore developed a new projection method that I will refer to as Ambient Embedding Projection (AEP). I had the idea to learn a multilayer perceptron $f_{\theta}(x_{ri}) : \mathbb{R}^d \rightarrow \mathbb{R}^3$ that projects a $d$-dimensional gene expression profile $x_{ri}$ of cell $i$ from biological replicate $r$ on to a two-dimensional surface embedded in three-dimensional Euclidean space. This enables visual inspection of both intrinsic and extrinsic geometry at the same time.

To estimate $\theta$, I asked Claude to suggest a loss function to measure the discrepancy between the intrinsic and extrinsic geometries of the projection and those of the original manifold, along with the departure of the projection surface from two dimensions. After many rounds of reading the literature it pointed me to, learning new material, playing with various alternatives, and implementing tests in Claude Code, I settled on the loss function
\begin{equation}
  \label{eq-aep}
  \sum_{r = 1}^n \left[
    \sum_{i = 1}^{m_r} \sum_{l = 1}^2 \frac{\{\sigma_l(J_{\theta,ri}T_{ri}) - 1\}^2}{2} +
  \frac{\lambda_{\mathrm{stress}}}{\sum_{i < j} \delta_{rij}^{2 - q}} \sum_{i<j} \delta_{rij}^{-q}(d_{\theta,rij}-\delta_{rij})^2 +
  \frac{\lambda_{\mathrm{rank}}}{m_r} \sum_{i = 1}^{m_r} \sigma_3(J_{\theta,ri})^2
  \right],
\end{equation}
where $n$ is the number of replicates in a given biological condition, $m_r$ is the number of cells from the $r$th replicate, $J_{\theta,ri}$ is the Jacobian $\partial f_\theta(x) / \partial x$ evaluated at $x_{ri}$, $T_{ri}$ is a $d \times 2$ orthonormal basis matrix of the tangent plane estimated at $x_{ri}$ using its $k$ nearest neighbors, $\sigma_l(\cdot)$ denotes the $l$th largest singular value of a matrix, and $\delta_{rij} = \Vert x_{ri} - x_{rj} \Vert$ and $d_{\theta,rij} = \Vert f_\theta(x_{ri}) - f_\theta(x_{rj}) \Vert$ are the Euclidean distances between cells $i$ and $j$ in the ambient and projected spaces, respectively.

The $k$, $q$, $\lambda_{\mathrm{stress}}$ and $\lambda_{\mathrm{rank}}$ are tuning parameters that I specified to be 20, 2, 5, and 1 by default, respectively. Results were fairly robust to these parameters, and I set $\lambda_{\mathrm{stress}}$ to be relatively large to make the extrinsic structure more apparent.

\begin{itemize}
\item The first term encourages the projection to match the manifold's intrinsic geometry. If the projection were an isometry, moving a small distance within the tangent plane of the manifold at $x_i$ should change $f_\theta(x_i)$ by the same distance, so the matrix $T_{ri}^\top J_{\theta,ri}^\top J_{\theta,ri} T_{ri}$ should equal the identity and the $\sigma_l(J_{\theta,ri}T_{ri})$ should equal 1. The first term therefore penalizes departures from isometry, and Claude designed it to match the scales of the other terms. It is a scaled version of the as-rigid-as-possible energy \citep{sorkine2007arap}.

\item The second term captures the manifold's extrinsic geometry by encouraging the Euclidean distances $\delta_{rij}$ in the original ambient space to be close to the Euclidean distances $d_{\theta,rij}$ in the projected space. The larger the tuning parameter $q$, the greater the influence of neighboring cells compared to cells far from each other. It is the same as stress terms used in multidimensional scaling and graph drawing \citep{chen2009local}. When there is more than one replicate, this term is calculated only with pairs of cells coming from the same replicate.

\item The third term penalizes the smallest singular value of $J_{\theta,ri}$ to encourage $f_\theta(x)$ to have a rank-2 Jacobian, so that the image of $f_\theta(x)$ a two-dimensional. An alternative approach would have been to specify one of the hidden layers of $f_\theta(x)$ to have only two nodes, so that the Jacobian is rank-2 by construction, but Claude noted that this would not be able to represent closed surfaces \citep{batson2021topological}.
\end{itemize}

When there is more than one biological replicate, \eqref{eq-aep} gives a sort of average projection across all replicates. An alternative approach might be to calculate different $f_\theta(x)$ for each replicate in a given condition and then take their Fr\'{e}chet mean.
  
I asked Claude to optimize $\theta$ using stochastic gradient descent. Briefly, it first precomputes neighbors and tangent bases in each replicate, then draws each minibatch from a single replicate chosen with probability proportional to the number of cells in the replicate. Claude found computationally efficient ways of calculating the loss so that visualizing around 35,000 cells from three replicates took less than a minute on a single CPU. I used Claude to summarize the details, which are provided in the Supplementary Material.

\begin{figure}
\centering{
\includegraphics[width=4.5in,height=\textheight]{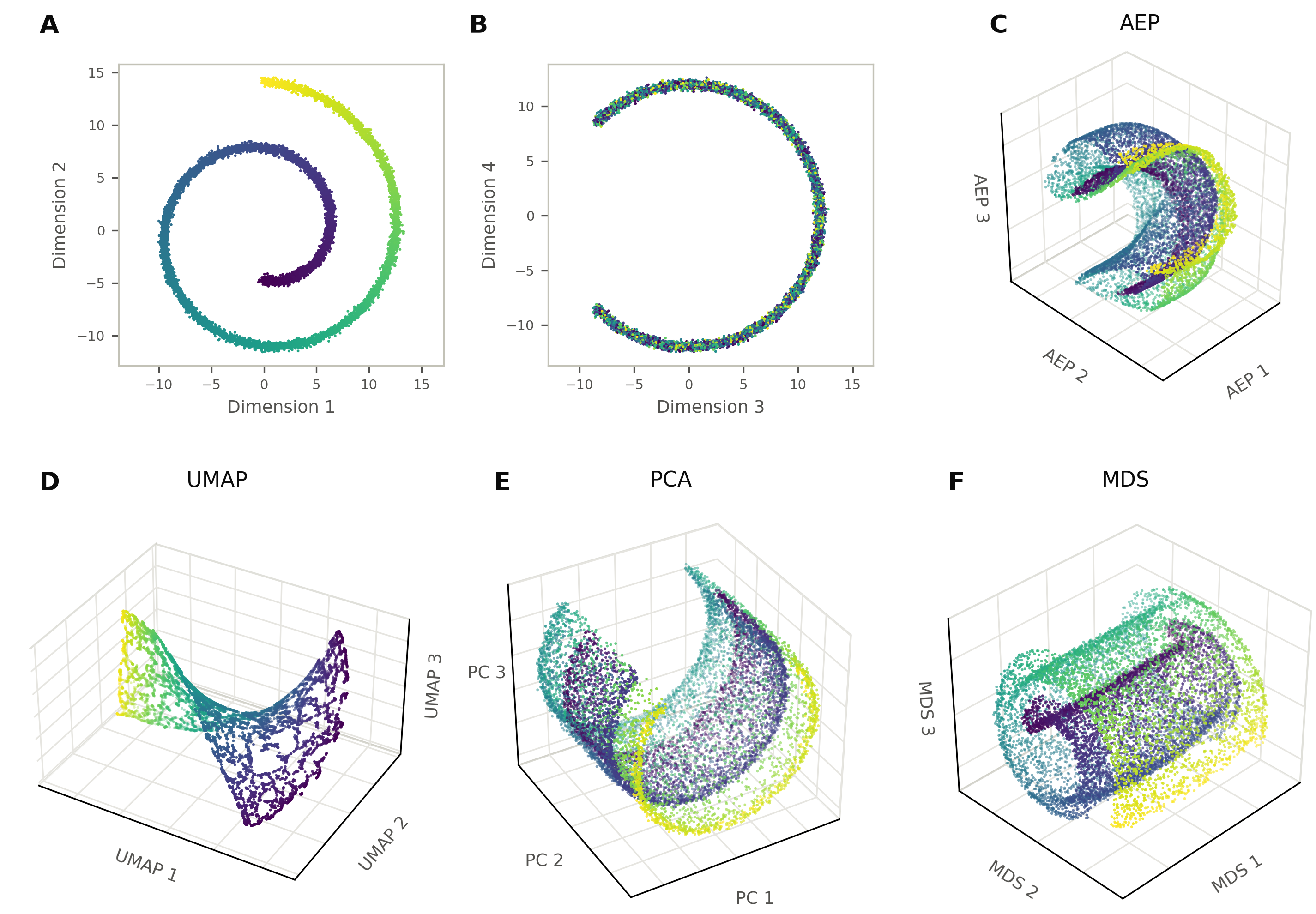}
}
\caption{\label{fig-swiss}Projections of 10,000 points from a four-dimensional swiss roll dataset. I used Claude Code to generate this figure.}

\end{figure}%

To test my new method, I first used Claude to generate a four-dimensional version of the swiss roll dataset. The first two dimensions (Figure \ref{fig-swiss}A) contained the roll. The height of the roll was governed by the third dimension, which changed as the fourth dimension changed (Figure \ref{fig-swiss}B). Imagine a video where time is the polar angle of the circle in Figure \ref{fig-swiss}B, so that as time progresses, the height of the swiss roll in the third dimensions goes up and down. A three-dimensional UMAP projection (Figure \ref{fig-swiss}D) captured the intrinsic two-dimensional nature of the data but did not show its extrinsic spiral structure. A PCA projection (Figure \ref{fig-swiss}E) captured the curvature in the third and fourth dimensions but not the spiral structure. An MDS projection calculated using the same stress as AEP (Figure \ref{fig-swiss}F) did not capture the curvature in the third and fourth dimensions. The AEP projection (Figure \ref{fig-swiss}C) reflected all of these features.

\subsection{Proposed explicational statistic}

I could now visualize the single-nuclei VTA data in Figure \ref{fig-vta} from \citet{minerva2026stress} using AEP. To ensure that the resulting projections would be meaningful across the control, resilient, and susceptible behavioral groups, I used Claude Science to generate batch-corrected gene expression values using scVI \citep{lopez2018deep} and identify the top 2000 most variable genes across replicates, following standard practice. Details are in the Supplementary Material. Within each biological condition, I then performed initial dimension reduction using the top 20 principal components of these highly variable genes, which explained more than 90\% of the variation. I did this separately wihin each condition to capture any features of the neurogenomic manifolds unique to the condition. I then used these to calculate the AEP projections.

\begin{figure}
\centering{
\includegraphics[width=6in,height=\textheight]{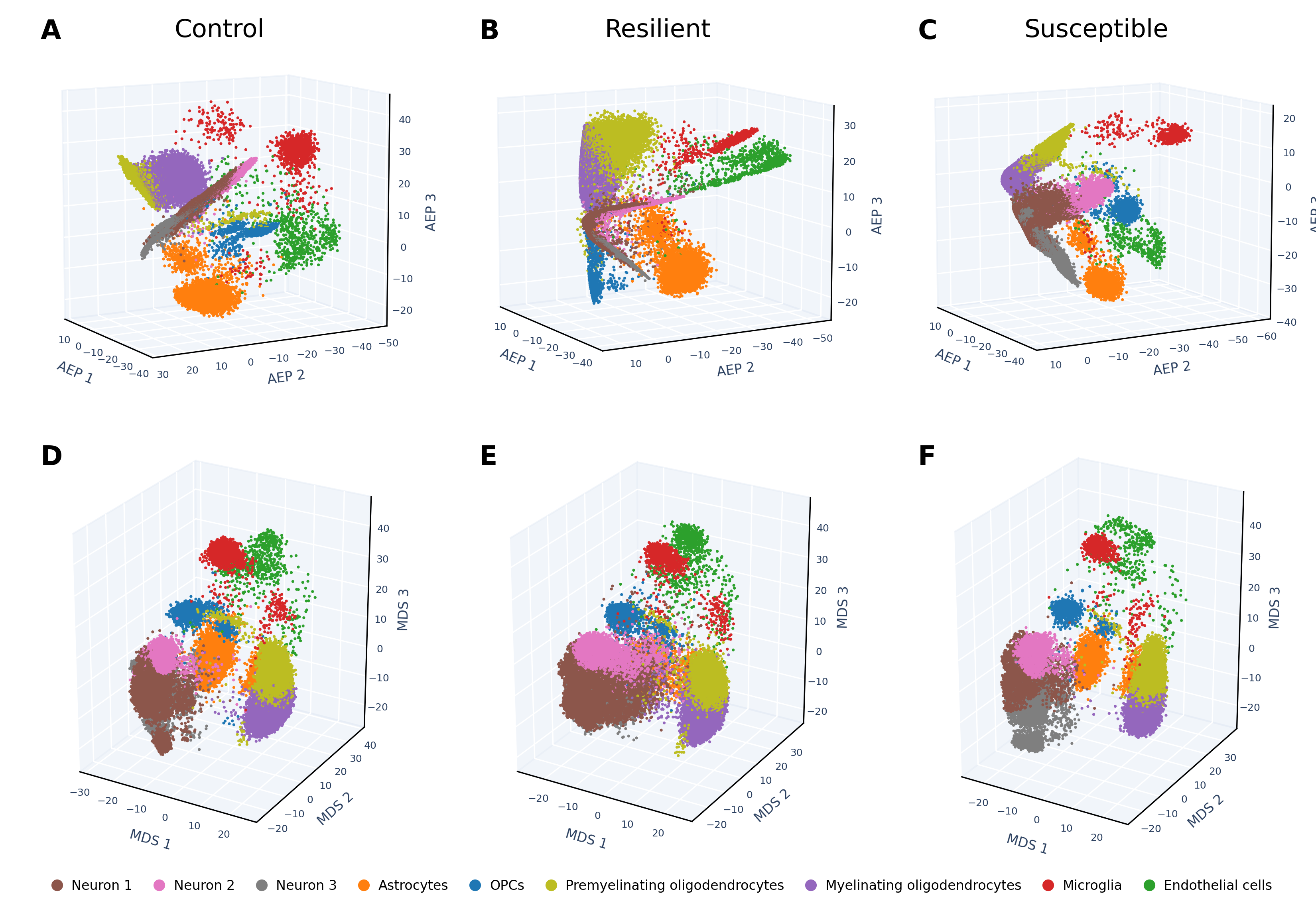}
}
\caption{\label{fig-vta_aep}Projections of snRNA-seq VTA data from \citet{minerva2026stress}. I used Claude Code to make this figure and an HTML file in the Supplementary Material where these projections can be interactively explored.}

\end{figure}%

Figures \ref{fig-vta_aep}A, \ref{fig-vta_aep}B, and \ref{fig-vta_aep}C show the AEP projections for each behavioral condition, fit with every replicate in each condition. They revealed several interesting changes in the shape and embedding of the neurogenomic manifolds that are not readily apparent from either the MDS projections made with the same stress as AEP (Figures \ref{fig-vta_aep}D, \ref{fig-vta_aep}E, and \ref{fig-vta_aep}F) or the UMAP projections in Figure \ref{fig-vta}. First, the neuronal cells appeared to have different extrinsic curvatures in the different behavioral conditions. Second, the intrinsic distances between the microglia and endothelial cells appeared to be somewhat similar in both the resilient and susceptible mice, but the extrinsic distances are dramatically different.

These observations suggest the following quantitative construct: let $X_{ri}$ and $X_{rj}$ be cells from replicate $r$ with cell types $C_{ri}$ and $C_{rj}$ and model them as lying on a neurogenomic manifold with an intrinsic distance metric $D_r(x,y)$ drawn from a distribution $\mathcal{D}$ on metrics. Since Euclidean distance will roughly equal the intrinsic distance at small scales and will always be larger at large scales, I model $\Vert X_{ri} - X_{rj} \Vert$ as a quadratic function of $D_r(X_{ri}, X_{rj})$, conditional on $D_r$ and the cell types:
\begin{equation}
  \label{eq-explication}
  \begin{aligned}
    &
      E_{\mathcal{D}} [E\{ \Vert X_{ri} - X_{rj} \Vert \mid  D_r = d_r, d_r(X_{ri}, X_{rj}) = h, C_{ri} = a, C_{rj} = b\}]\\
    =\,&
         \beta_{0ab} + \beta_{1ab} h + \beta_{2ab} h^2.
  \end{aligned}
\end{equation}

\begin{figure}
\centering{
\includegraphics[width=2.5in,height=\textheight]{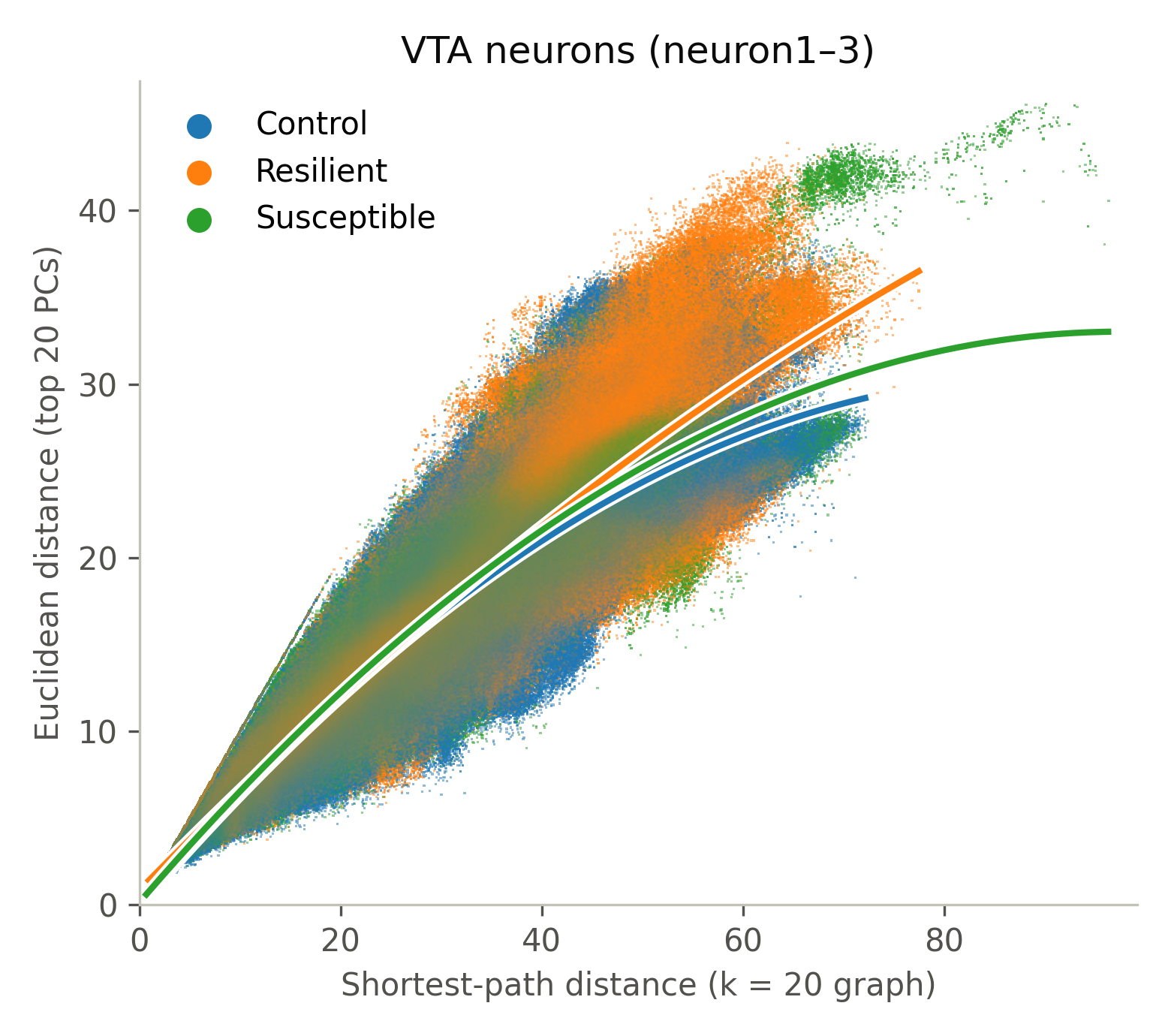}
\includegraphics[width=2.5in,height=\textheight]{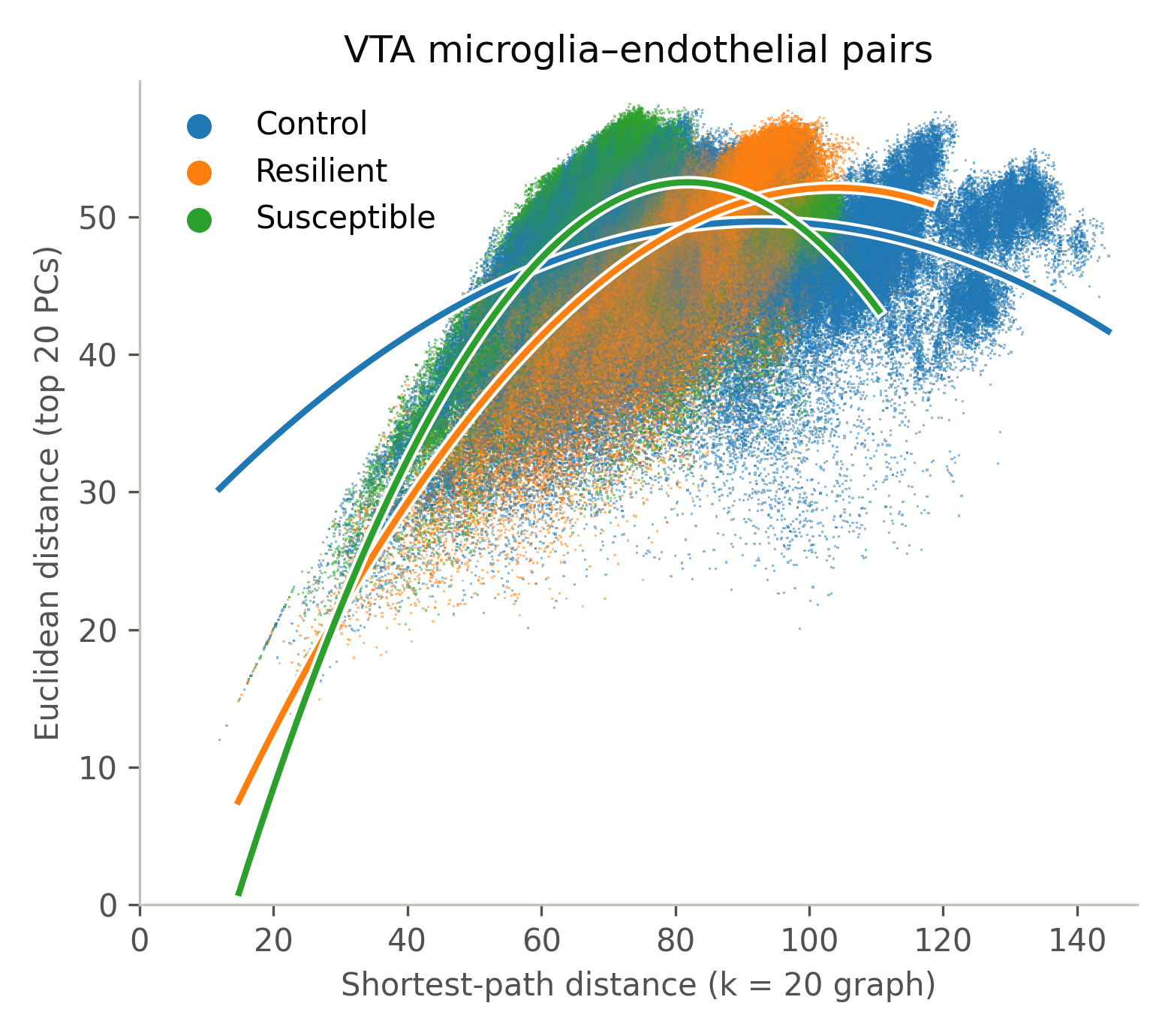}
}
\caption{\label{fig-b2a}Proposed $\beta_{2ab}$ \eqref{eq-explication} explicates transcriptomic convergence in cell biology. Fitting \eqref{eq-explication} using least-squares (solid lines) reveals distinct variation in $\beta_{2ab}$ across behavioral conditions. I generated this figure with Claude Code.}

\end{figure}

Based on this model, I propose $\beta_{2ab}$ as a new explication of an ensemble property of cellular transcriptomes, which might be thought of \textit{transcriptomic convergence}. Intuitively, the more negative it is, the more we will find that cells of type $a$ are intrinsically far from cells of type $b$, yet have similar gene expression values. This might suggest that $a$ and $b$ have taken on complementary functional roles within the overall cellular ensemble. This is an ensemble property because it cannot be calculated from the two cell types alone and depends on the overall structure of the neurogenomic manifold. Figure \ref{fig-b2a} illustrates initial least-squares fits of \eqref{eq-explication} using all pairs of cells in neurons 1, 2, and 3 and all microglia-endothelial cell pairs that show distinct variation between behavioral groups.

My $\beta_{2ab}$ construct appears not to have been studied yet in the literature. However, using Claude I discovered that it is related to deep areas of biology, mathematics, and statistics. Biologically, phenotypic convergence has been observed in distinct neuronal cell types \citep{konstantinides2018phenotypic}. Mathematically, $\beta_{2ab}$ is related to the notion of distortion introduced by \citet{gromov1992asymptotic} and used in the study of extrinsic geometry \citep{farb1994extrinsic, mitra1998coarse}. Statistically, there is recent statistical work on estimating the reach of a manifold \citep{aamari2023optimal, cholaquidis2024notion}. My new explication thus opens the door to rich new programs of research, and may be of broader interest beyond cell biology.

\section{Discussion}\label{sec-discussion}

In this paper I have aimed to delineate explicational statistics as a distinct category of statistical knowledge. To give a concrete illustration, I developed a new dimension reduction method, Ambient Embedding Projection \eqref{eq-aep}, and used it to conceive of a new quantitative construct, $\beta_{2ab}$ \eqref{eq-explication}, to explicate the putative theoretical construct of transcriptomic convergence, an ensemble property of single-cell transcriptomes.

Several aspects of this illustration echo my arguments in Section \ref{sec-ai} about why research in explicational statistics should become increasingly valuable in the age of AI. First, AI is a powerful tool that is most useful when pointed at precise quantitative problems. New explications provide these problems. My definition of $\beta_{2ab}$ raises numerous new issues of optimal estimation, inference, and application, which AI can be deployed to resolve.

Second, I could not have used AI to automate the explication process, because I did not start out knowing the right biological construct to explicate. I developed the ideas of transcriptomic convergence and $\beta_{2ab}$ simultaneously, and each helped motivate the other. Visualization is a useful tool for conceptual co-construction \citep{tukey1975mathematics}, and was integral in the proposal of $\beta_{2ab}$ and in the thought experiment in Section \ref{sec-research} with the bimodal American age distribution.

Finally, my ultimate choice of $\beta_{2ab}$ as the right quantitive construct was driven in part by my judgement of the value structures of the fields of biology and statistics. My judgement is not replicated by AI. As an experiment, I asked Claude to suggest, on its own, the best quantitative construct to capture the ensemble behavior of cellular transcriptomes. It gave an answer, which is included in the Supplementary Material, that looks very different from mine. For example, its alternative approach explicates a different biological construct and is not related to manifold learning, whereas I felt that a manifold-based construct would be interesting to statisticians, computer scientists, and neuroscientists familiar with neural manifolds.

My delineation of explicational statistics speaks to what I think is a larger trend in the production of knowledge. In his account of the philosophy of design, \citet{parsons2015philosophy} argued that the industrial revolution led to the separation of the design of material goods from their production. Historically, both were carried out by a single craftsman, but industrialization turned them into specialized tasks now carried out by distinct groups. Something similar seems to be playing out in the modern industrialization of knowledge production. Explication may now become a separate task as AI increasingly automates other parts of the statistical research process. This suggests that researchers might need to be trained in part like architects or product designers, who need to understand the processes of production but do not carry out all of it themselves.

\section{Data Availability Statement}\label{data-availability-statement}

The snRNA-seq data are publicly available from the Gene Expression Omnibus (GSE303538). The AEP implementation is available at \url{https://github.com/sdzhao/aep/}.

\phantomsection\label{supplementary-material}
\bigskip

\begin{center}

{\large\bf SUPPLEMENTARY MATERIAL}

\end{center}

\begin{description}
\item[aep\_supplement.pdf:]
  Detailed description of the AEP method, generated by Claude Code.
\item[Alternative explication - Claude.pdf:]
  Alternative explication of cellular ensemble properties identifed autonomously by Claude.
\item[GSE303538\_methods.pdf:]
  Detailed description of the processing of the snRNA-seq data, generated by Claude Science.
\item[GSE303538\_pipeline.py:]
  Python script for processing the VTA snRNA-seq data, generated by Claude Science.
\item[results.html:]
  Interactive projections of the VTA snRNA-seq data, generated by Claude Code.
\end{description}

\bibliography{bibliography.bib}

\end{document}